\documentclass{article}

\usepackage{color}
\usepackage{times}
\usepackage{epsfig}
\usepackage{graphicx}
\usepackage{amssymb}
\usepackage{amsmath}
\usepackage{amssymb}
\usepackage{url}
\usepackage{stfloats}
\usepackage{placeins}
\usepackage{caption}
\usepackage{multicol}
\usepackage[margin=1.1in]{geometry}
 
\title{An Automated Images-to-Graphs Framework for High Resolution Connectomics}

\author{William Gray Roncal,$^{1,2,*}$, Dean M Kleissas,$^{2,*}$, Joshua T Vogelstein $^{3,4}$, Priya Manavalan $^{1}$,\\  Kunal Lillaney $^{1}$, Michael Pekala $^{2}$, Randal Burns $^{1}$, R Jacob Vogelstein $^{2}$,\\ Carey E Priebe $^{5}$, Mark A Chevillet $^{2}$, Gregory D Hager,$^{1}$ 
\\ 
\\
$^{1}$Johns Hopkins University, Department of Computer Science, Baltimore, Maryland\\
$^{2}$JHU Applied Physics Laboratory, Research and Exploratory Development, Laurel, Maryland \\
$^{3}$Johns Hopkins University, Department of Biomedical Engineering, Baltimore, Maryland\\
$^{4}$Johns Hopkins University, Institute for Computational Medicine, Baltimore, Maryland \\
$^{5}$Johns Hopkins University, Department of Applied Math and Statistics, Baltimore, Maryland \\
Corresponding Author:  William Gray Roncal, wgr@jhu.edu \\
$^{*}$ Authors contributed equally \\ 
\rule{3cm}{0.05cm}\\
}

\begin{document}

\maketitle

\begin{abstract}

Reconstructing a map of neuronal connectivity is a critical challenge in contemporary neuroscience.  Recent advances in high-throughput serial section electron microscopy (EM) have produced massive 3D image volumes of nanoscale brain tissue for the first time.  The resolution of EM allows for individual neurons and their synaptic connections to be directly observed.  Recovering neuronal networks by manually tracing each neuronal process at this scale is unmanageable, and therefore researchers are developing automated image processing modules.  Thus far, state-of-the-art algorithms focus only on the solution to a particular task (e.g., neuron segmentation or synapse identification). 

In this manuscript we present the first fully automated images-to-graphs pipeline (i.e., a pipeline that begins with an imaged volume of neural tissue and produces a brain graph without any human interaction).  To evaluate overall performance and select the best parameters and methods, we also develop a metric to assess the quality of the output graphs.  We evaluate a set of algorithms and parameters, searching possible operating points to identify the best available brain graph for our assessment metric.  Finally, we deploy a reference end-to-end version of the pipeline on a large, publicly available data set.  This provides a baseline result and framework for community analysis and future algorithm development and testing.  All code and data derivatives have been made publicly available toward eventually unlocking new biofidelic computational primitives and understanding of neuropathologies.

\vspace{0.5cm}

\end{abstract}

\section{Introduction}

Brain tissue volumes imaged using electron microscopy today already contain many thousands of cells that can be resolved at the scale of a single synapse. The amount of information is daunting: in just $1 \, mm^3$ of brain tissue, we expect petabytes of data containing $10^5$ neurons and $10^9$ synapses \cite{bs1991}.  While this region is very small in physical volume compared to an entire brain, it is roughly the scale of a cortical column, a hypothetical fundamental organizing structure in the cortex \cite{Mountcastle1957}.

\begin{figure}[h]
\begin{center}
\includegraphics[width=8.5cm]{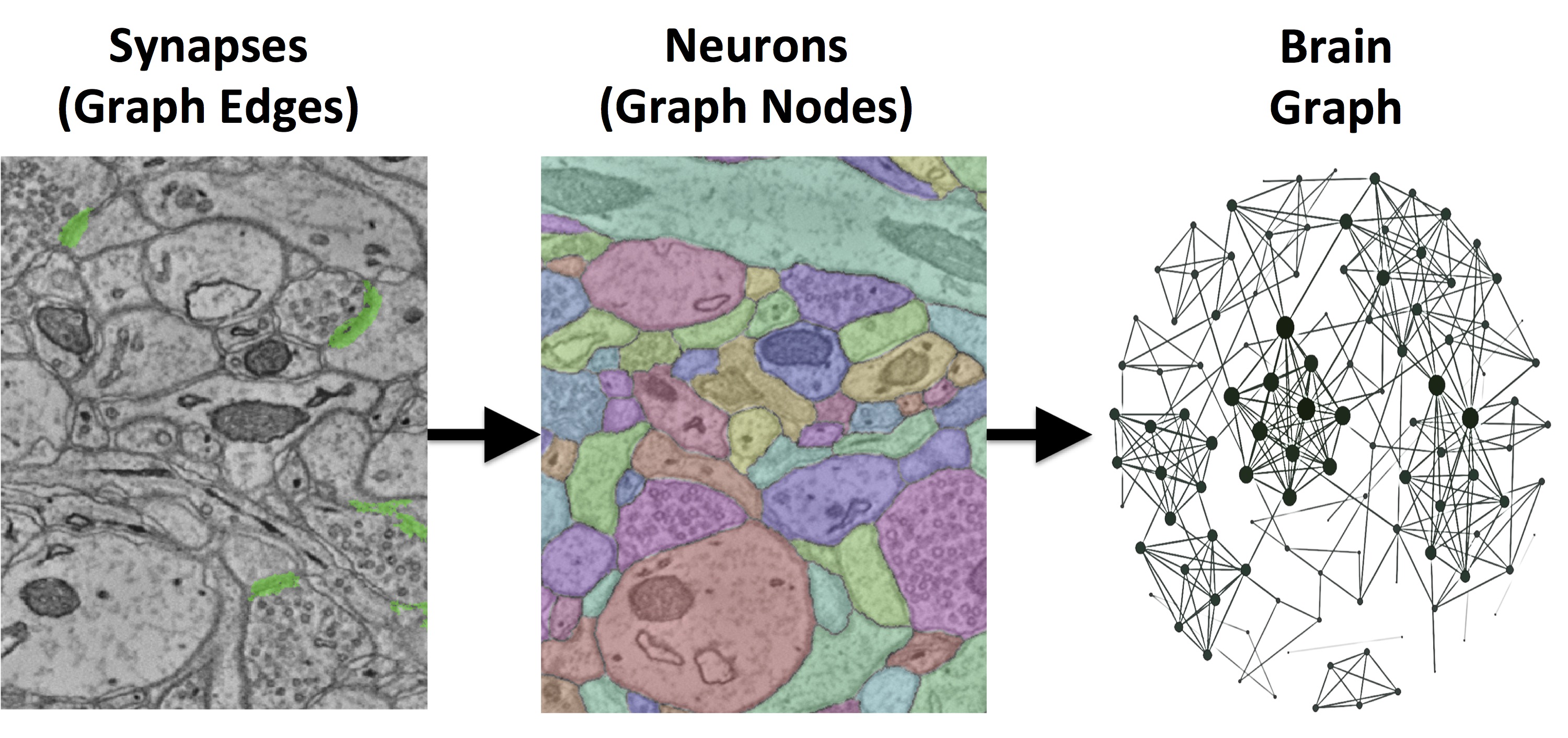}
\caption{An illustration of the images-to-graphs process.  Detected synapses are shown in green (left); these are combined with multicolored neuron segments (middle) into a graph (right). Nodes are represented by neurons and edges by synapses.}
\label{i2g}
\end{center}
\end{figure}

Our goal is to transform large 3D electron microscopy volumes of neural tissue into a detailed connectivity map, called a connectome (Figure~\ref{i2g}). This approach will directly estimate brain graphs at an unprecedented level of detail.  Each neuron is represented in the graph as a node, and each synapse is represented as an edge connecting these nodes.  Manual human annotation, while currently the most accurate method of reconstruction is unrealistic as volumes scale.  A recent study estimated that manual annotation of a cortical column requires hundreds of thousands of person-years \cite{Helmstaedter2011a}.

Therefore, an automated method to run algorithms at scale is needed to realize the promise of large-scale brain maps.  We developed a novel ecosystem of algorithms, software tools and web services to enable the efficient execution of large-scale computer vision algorithms to accomplish this task.

We also introduce a fully automated images-to-graphs pipeline and an assessment metric for the resulting graphs.  This metric allows us to directly assess the connectivity properties of the graph, rather than relying on intermediate measures (e.g., synapse precision-recall or segmentation pixel error).    We run a grid search over a collection of parameters (i.e., both individual modules and their settings) using our pipeline to determine the best available result for analysis and interpretation.  Once this optimal operating point was determined, we estimated the brain graph for a volume of neural tissue in our scalable framework.

\subsection{Previous Work}
Previous research has produced methods that advance the field of connectomics in important ways, but none have provided an end-to-end, automated, scalable approach. Several manual or semi-automated approaches have been used to construct brain circuits \cite{Bock2011, Takemura2013, Mishchenko2010}. 
Other groups have produced automated algorithms \cite{Kaynig2013b,Sommer2011,Nunez-Iglesias2013} that solve important pieces of the overall puzzle (e.g., neuron segmentation, synapse detection).  These modules have generally been evaluated on small subvolumes without considering the overall graph result;  additional work is needed to improve both algorithm accuracy and scalability for large graph inference.  

In building our images-to-graphs pipeline, we leveraged previous work whenever available.  To detect cell membranes, we reimplemented the ISBI 2012 challenge-winning approach \cite{ciresan2012deep}, which framed membrane detection as a per-pixel classification problem and obtained state-of-the-art results using a small ensemble of Deep Neural Networks (DNN). 
We segment the neuronal structures by incorporating Rhoana \cite{Kaynig2013b}, an open-source algorithm, which selects and fuses candidate 2D contours into 3D objects using conditional random fields (CRFs) and Integer Linear Programming (ILP).  We also integrate Gala \cite{Nunez-Iglesias2013}, an agglomerative approach that combines super pixels into neurons.  Together these two methods represent the two major approaches currently used in neuron segmentation; other methods can be readily adapted to this framework if desired.  Scalable synapse detection (i.e., the edges in our graph) is still an open problem.  While partial solutions exist \cite{Kreshuk2014, Becker2012a}, they were developed for specific imaging paradigms (e.g., isotropic, post-stained).  Therefore, we developed our own lightweight, scalable synapse detector.
Finally, the Open Connectome Project \cite{Burns2013} provides a high-performance spatial database optimized for processing large neuroimaging volumes.  These tools facilitate scalable processing and provide significant advances over a flat-file architecture in terms of data storage and access.

\section{Framework}

Our tools are built on a distributed processing framework which leverages the LONI Pipeline  \cite{Rex2003a} for workflow management, and interfaces with the data and annotation services provided by the Open Connectome Project (OCP)  \url{openconnecto.me} .  It includes an application programming interface (API) which implements our data standard for high resolution connectomics, and tools to enable rapid algorithm development while simplifying the challenge of running at scale.  Our framework enables researchers to focus on developing novel image processing techniques without considering data management and processing challenges.

We are able to efficiently incorporate new methods by extracting only core algorithm code (often a small fraction of the original code base).  We reuse our data management framework, eliminating the need for researchers to rewrite solutions for file handling and bookkeeping. This capability enables image processing researchers to build state-of-the-art algorithms without addressing cluster integration or scalability details.

\subsection{RAMON}

An acknowledged challenge in the connectomics field is annotation representation and its impact on software and institution-level interoperability.  As the field grows and data volumes increase, the sharing of data through remote and programmatic interfaces and the application of community developed algorithms and software will become common.

Answering this challenge requires scene parsing, rather than simply segmentation; the rich semantic annotations are critical to inferring graph structure and understanding the function and structure of neuronal circuits. We developed a standard for annotation metadata, as summarized in Table~\ref{table:RAMON}, which we call the Reusable Annotation Markup for Open coNnectomes (RAMON).  

RAMON defines a minimum set of annotation types and associated metadata that capture important biological information and build the relationships between annotations that are critical for connectome generation. Annotation metadata is trivially extensible through custom, user defined key-value pairs.  This is not a formal ontology; rather it facilitates the development of software and tools by providing a flexible, yet reliable, standard for representing annotation data. For example, our synapse annotation type has fields such as weight, type, confidence, associated neural segments, and is extensible with arbitrary, searchable key-value pairs. 

\begin{table}[h!]
 \begin{center}
  \resizebox{12cm}{!} {

  \begin{tabular}{|p{3cm}|p{9cm}|}

  \hline
  \textbf{Type} & \textbf{Description}  \\\hline
  SYNAPSE & Junction between two NEURONs is used to connect SEGMENTs when building a GRAPH \\ \hline
  ORGANELLE &  Represents internal cell structures (e.g, mitochondria, vesicles) \\ \hline
   SEGMENT & A labeled region of a neuron; typically a contiguous voxel set  \\ \hline
  NEURON & Container for assembling SEGMENTs \\ \hline
  VOLUME & Used to store pixel label information; inherited by many other types \\ \hline
  GENERIC & Extensible, used to specify arbitrary, user-defined information for a voxel set \\ \hline

 \end{tabular}
  }
    \end{center}

    \label{table:RAMON}

An overview of the current RAMON object types.  Each object is used to provide labels and attributes to objects identified in the neural tissue.  This facilitates interoperability and efficient data storage, retrieval, and query processing.
\end{table}

\subsection{OCP}

Our approach for creating brain graphs from large image volumes (Images-to-Graphs) leverages the Open Connectome Project (\url{www.openconnecto.me}) infrastructure \cite{Burns2013}. 

Open Connectome Project (OCP) is a distributed, scalable cluster, designed for spatial analysis and annotation of high-resolution brain data. It supports data from multiple imaging modalities (e.g., electron microscopy, light microscopy, array tomography, time series) with the goal of providing a unified backend for connectomics research. The system is designed using NoSQL data-stores and data-intensive architectures. The spatial index is partitioned, thus distributing the data across all the data-nodes to maximize throughput and avoid I/O interference.

\begin{figure*}[tb]
\includegraphics[width=\textwidth]{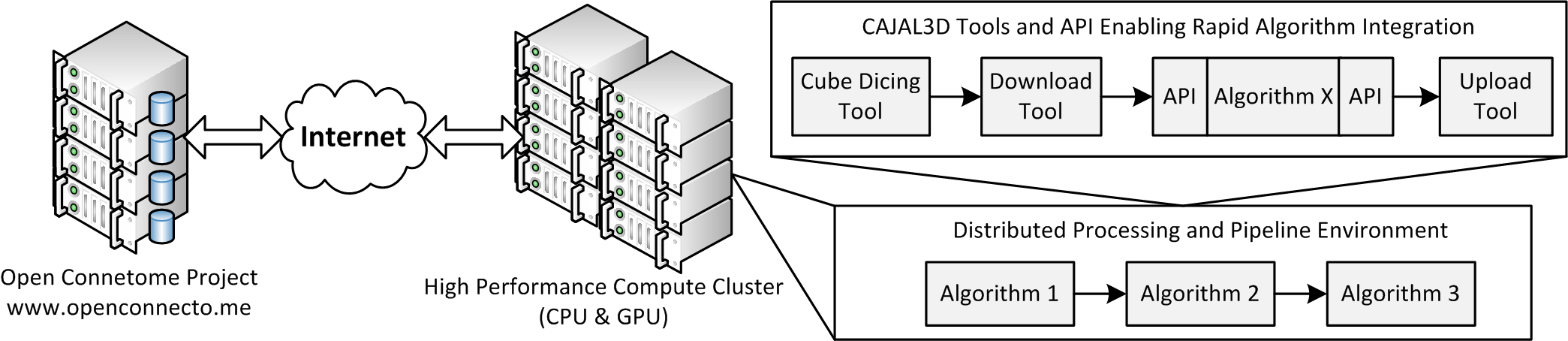}
\caption{An overall view of the processing framework, illustrating a distributed processing framework.  Data and annotation stores leverage the Open Connectome Project, and interface with a high perforamnce compute cluster.  A variety of tools are available to facilitate a distributed processing environment.}
\label{ocppipeline}
\end{figure*}

OCP implements RAMON and provides publicly-accessible RESTful (REpresentational State Transfer) web-services that enable efficient data storage, retrieval and queries for images-to-graphs.  This infrastructure provides the capability to efficiently view arbitrary data planes and access data subvolumes.  Our pipelines store the output of computer vision algorithms as a set of queryable label and metadata annotations in an OCP spatial database co-registered with the image dataset. OCP currently provides a unified interface and backend for connectomics and stores more than 70 unique data sets and thousands of annotation projects totaling more than 50 TB of neural imaging data on disk; these data encompass many different modalities and scales.  

\subsection{Application Programming Interface}
To enable rapid software development, we have created  Connectome Annotation for Joint Analysis of Large data (CAJAL), an open source MATLAB Application Programming Interface (API), which implements the RAMON standard.  The API provides classes for all RAMON annotation types, allowing users to easily develop interoperable algorithms and analysis tools. In addition to storing label data and metadata, the RAMON classes provide additional functionality such as tracking the annotation's global database location and visualizing a region of interest.

The API also provides an easy to use interface to query, upload, and download data from OCP services that abstracts complicated RESTful URL generation from the user.  Many data formats are supported for both image and annotation data, including grayscale and multi-channel image data and integer and floating point based annotations. The toolbox automatically handles compression, chunking, and batching of annotation data to optimize throughput and simplify software development. 

\subsection{Infrastructure}

Our infrastructure is divided into front end processing and back-end Open Connectome Project Services as shown in Figure~\ref{ocppipeline}.

\subsubsection{Front-End}

Data and results were stored using the Open Connectome Project.  Our CPU compute cluster for images-to-graphs evaluation and deployment used 100 AMD Opteron 6348 cores (2.8GHz, 12-core/processor, 4 processors/node) and 256GB of RAM per node ($\approx$ 1TB total).  For membrane detection, we use a small GPU cluster containing 27 GeForce GTX Titan cards with 6GB RAM.   We leverage Son of Grid Engine (SGE) for task scheduling and the LONI Pipeline for workflow management \cite{Rex2003a}. Because we parallelize at a data block level, each task is Embarassingly Parallel, and so we use traditional scheduling methods. 

\subsubsection{Back-End}
On the backend, OCP uses a load-balancing webserver (2x Intel Xeon X5650 2.67GHz 12 core/processor and 48 GB of RAM).  This webserver distributes jobs across three data servers running a distributed database (each with 2x Intel Xeon X5690 3.47GHz 12 core/processor and 48GB of RAM).  Additionally, 100TB of network mounted disk-space is available for storage \cite{Burns2013}.

\subsection{Distributed Block Processing}
\begin{figure}[h]
\begin{center}
\includegraphics[width=7cm]{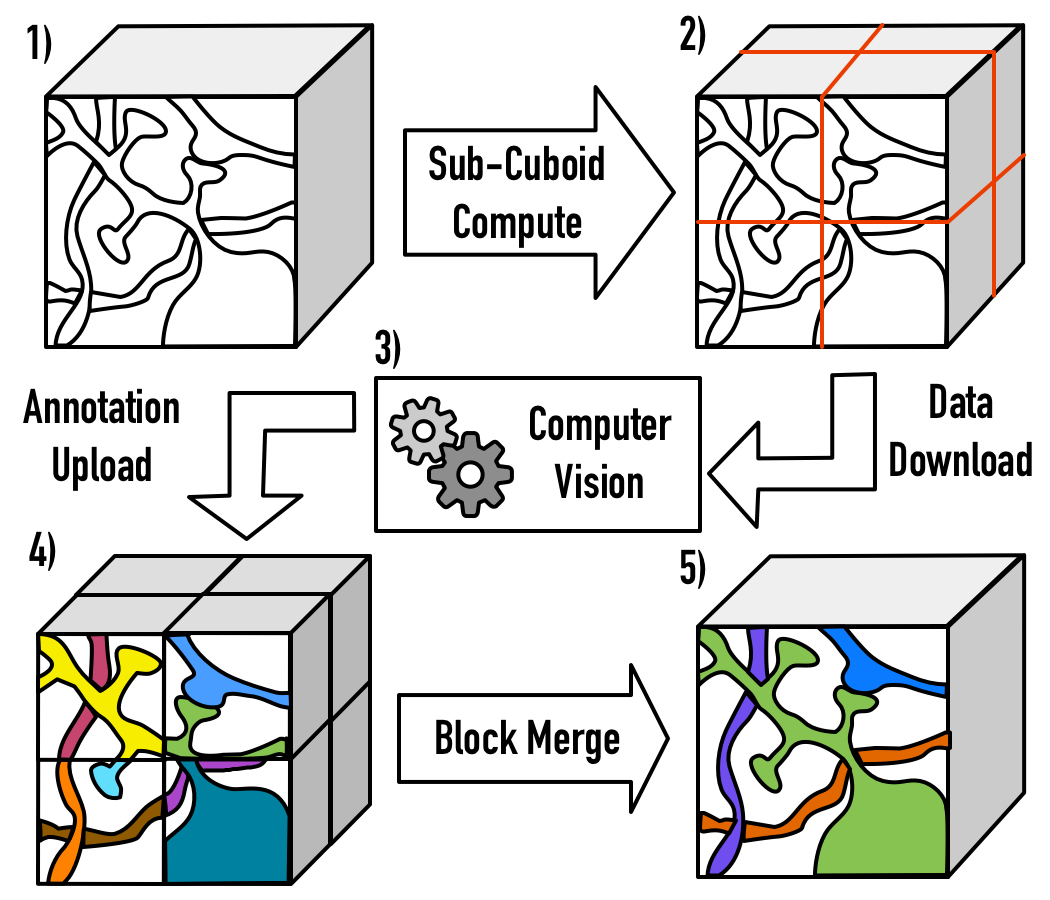}
\caption{An illustration of the distributed processing paradigm. Raw image data (1) is divided into cuboids (2) based on user-specified parameters , with the necessary padding to perform computation.  After processing (3), the annotations are inscribed and uploaded to OCP (4).  Finally, processes are merged across block seams (5), using a similarity metric of the user\'s choice.}
\label{blockmerge}
\end{center}
\end{figure}

One major open challenge (e.g., in a file-based image system) is scaling Vision algorithms to massive volumes.  We resolve this by extracting cuboids from large data volumes, accounting for overlapped and inscribed regions; processing the block; and then merging the annotations across the resulting cuboid boundaries.  The current implementation serially processes large block seams, eliminating the need for transitive closure operations.  As volumes scale, a hierarchical approach may be desirable to increase throughput (Figure~\ref{blockmerge}).  

\section{Images-to-Graphs Pipeline}
Often computer vision and other analytic tools for connectomics are written targeting a specific dataset, analysis product, or lab, and are only run on a limited data volume.  Also, researchers invest much time in developing the required support software to manage data and facilitate processing.  This results in code that is challenging to port and results that can be difficult to share or reproduce.  To leverage state of the art ideas across the connectomics community, a flexible and robust system is beneficial to integrate and evaluate disparate tools, especially when scaling  processing to massive data volumes.

We set out to build such a framework for connectomics processing that was agnostic to the underlying algorithms and provided reusable modules for common steps such as volume dicing, image data download, annotation upload, and annotation merging. By leveraging RAMON, Open Connectome Project, our API, and the LONI Pipeline workflow manager \cite{Rex2003a}, we built a system capable of rapidly integrating, running, and evaluating connecomics algorithms on a single workstation or on a high performance compute cluster.

\subsection{Assessment Measures}

A variety of error measures have been proposed for connectomics (e.g.,  warping error, adjusted Rand index, variation of information \cite{Nunez-Iglesias2013}), but are limited by their focus on an individual subtask of the entire images-to-graphs pipeline.  As we demonstrate, the optimal results for a subtask may not translate to optimal results for the overall pipeline.  

\begin{figure}[htbp]
\begin{center}
\includegraphics[width=8.5cm]{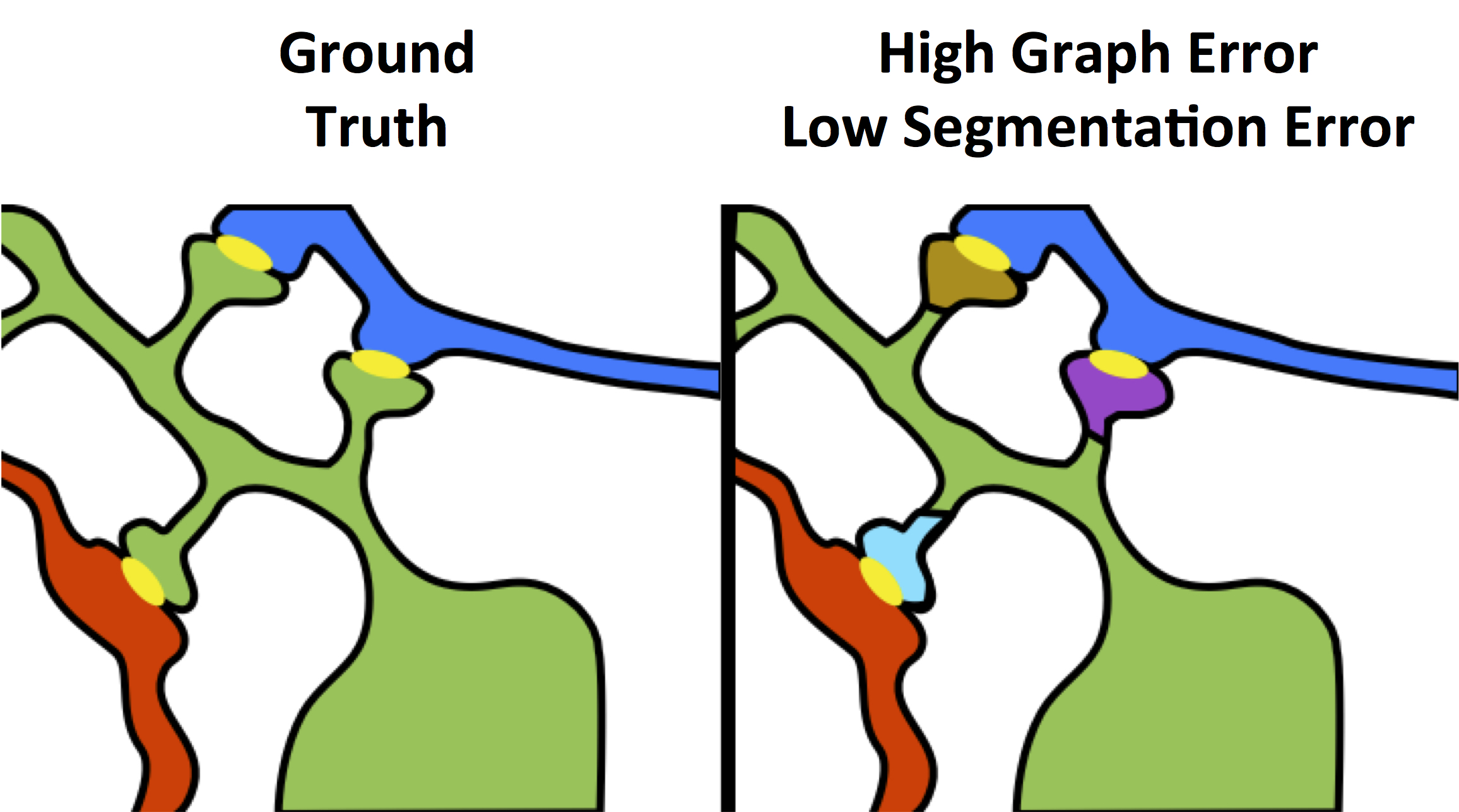}
\caption{Yellow objects represent synaptic junctions; other colors are different neurons.  The left panel shows true connectivity; the right panel shows the effect of fragmenting neurons at the dendritic spine necks, which produces a very small change in segmentation error, but a dramatic impact to graph error.}  \label{grapherror}
\end{center}
\end{figure}

As shown in Figure~\ref{grapherror}, even small neuron segmentation errors that affect graph connections are potentially very significant in terms of the resulting graph, while large errors not affecting topology may be much less significant.  These small, significant errors occur frequently on narrow, fragmented spine necks, breaking connectivity between the synapse on the spine head and the parent dendritic shaft \cite{Kaynig2013b}.

\begin{figure}[h]
\begin{center}
\includegraphics[width=8.5cm]{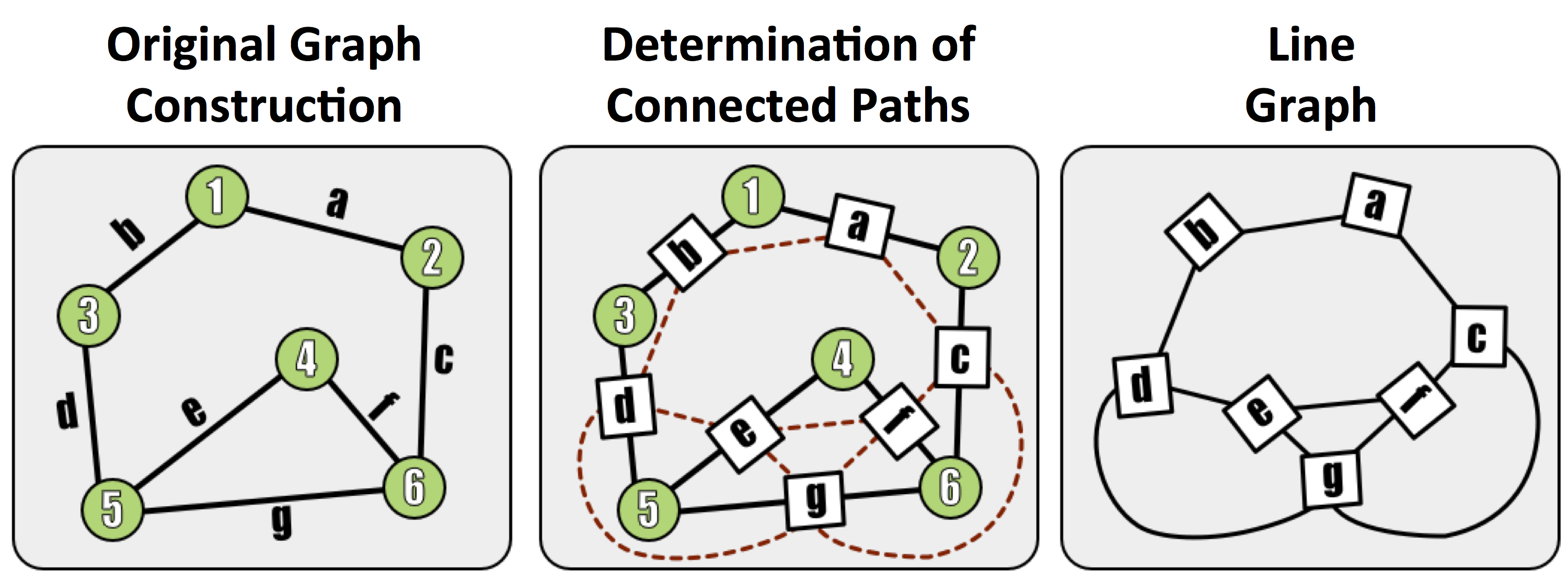}
\caption{A demonstration of the methods used to construct a line (edge-based) graph from a conventional node based network, by finding paths between edges in the original graph.}
\label{linegraph}
\end{center}
\end{figure}

To assess graph error, we first form the \textit{line graph}, which is the dual of the traditional graph, and represents connections (i.e., paths) between terminals.  In this formulation, the synapses become the graph nodes and the graph edges are constructed from the neurons (Figure~\ref{linegraph}).  

Line graphs are constructed for both the estimated \\ $\mathcal{L}\{G_{estimated}\}$ and true $\mathcal{L}\{G_{true}\}$ neuronal graphs, resulting in square, undirected, binary upper triangular matrices.  To directly compare the graphs, we augment both matrices so that every node (i.e., synapse) in both graphs has a direct correspondence.  We first find synapses in common by overlapping synapse annotations, and then add synapses absent in one graph, but present in the other, to both the true and test graphs. This graph correspondence is much easier to determine in the line graph (since synapses are small, compact objects) than in the traditional graph formulation (which often contains many neuron fragments).  

We propose two graph error metrics; the first is simply the Frobenius norm between the true and test graphs: (Eqn~\ref{lgeqn}).

\vspace{-0.25cm}
\begin{equation}
\label{lgeqn}
G_{err} = || \mathcal{L}\{G^*_{true}\} - \mathcal{L}\{G^*_{estimated}\}||_{F}
\end{equation}

This metric is attractive in its simplicity, but has a few major disadvantages.  This measure is unbounded, and the error will tend to increase with graph size; it is potentially misleading because it rewards the disproportionately large number of true negative edges in sparse graphs.

The second formulation computes the f1 score of the detected edges in the line graph, compared to a truth baseline \ref{lgeqn2}.  In this paradigm, true positive edges (TP) are occur in both the true and test graphs; false positive edges (FP) are present in the test graph and not in the true graph, and false negative edges (FN) are true edges missed in the test graph.  Similar to an object detection setting, precision, recall, and the f1 score are computed for the test graph:

\noindent\begin{minipage}{.5\linewidth}
\begin{equation}
    Precision = \frac{TP}{TP + FP}
\end{equation}
\end{minipage}%
\begin{minipage}{.5\linewidth}
\begin{equation}
    Recall = \frac{TP}{TP + FN}
\end{equation}
\end{minipage}

 \begin{equation}
    f1_{graph} = \frac{2 \times Precision \times Recall}{Precision + Recall}
\end{equation}
  \label{lgeqn2}

Our metric is interpretable, because true connections are the non-zero common entries. Furthermore, each incorrect entry represents a false positive (spurious connection) or false negative (missed connection).  A connection between two synapses in a line graph is equivalent to those synapses being coincident on a neuron.  This metric has scalability advantages over voxel-based metrics, because it can be easily computed on large volumes and can be used to characterize common errors.  This measure is on [0,1], and is robust to graph sparseness - true negatives do not impact the metric, and this prevents a class of misleading results (e.g., an empty graph).

For our application (and in this manuscript) we optimize algorithm selection based on graph f1 score.  Researchers may choose a different optimation goal depending on their application (e.g., maximizing recall with acceptably high precision).  A variety of metrics are computed (TP, FP, FN, precision, recall, f1, Frobenius norm) for each graph and available online.

\subsection{Algorithms}

Our approach transforms an image volume of cortical tissue into a wiring diagram of the brain.  To assemble this pipeline, we begin with membrane detection \cite{ciresan2012deep}, and then assemble these putative two-dimensional neuron segments into three-dimensional neuron segments, using  Rhoana \cite{Kaynig2013b}, Gala \cite{Nunez-Iglesias2013}, or a watershed-based approach.  These are the nodes in our graph, and are compared using the Adjusted Rand Index, computed as a comparison to neuroanatomist created ground truth, following community convention. 

\begin{figure}[h]
\begin{center}
\includegraphics[width=8cm]{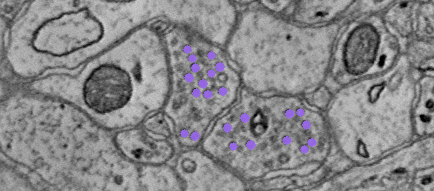}
\caption{Depiction of neurotransmitter-containing vesicles.  The presence of synaptic vesicles is the most important feature for our Random Forest, and likely for many manual annotators.} 
\label{vesicles}
\end{center}
\end{figure}

To find the graph edges, we develop a lightweight, scalable Random Forest synapse classifier.  We combine texture (i.e., intensity measures, image gradient magnitude, local binary patterns, structure tensors) with biological context (i.e., membranes and vesicles). We pruned a large feature set to just ten features and used this information to train a random forest.  The pixel-level probabilities produced by the classifier were grouped into objects using size and probability thresholds and a connected component analysis.   This method requires substantially less RAM than previous methods, which enables large-scale processing.   A key insight was identifying neurotransmitter-containing vesicles present near (chemical, mammalian) synapses.  These were located using a lightweight template correlation method and clustering (Figure~\ref{vesicles}). 

Performance was further enhanced by leveraging the high probability membrane voxels (described above), to restrict our search, improving both speed and precision.  Our synapse performance was evaluated using the F1 object detection score, computed as the harmonic mean of the precision and recall, based on a  gold-standard, neuroanatomist-derived dataset.  We took care to define our object detection metric to disallow a single detection counting for multiple true synapses, as that result is ambiguous and allows for a single detection covering the whole volume to produce an F1 score of 1 \cite{GrayRoncal}.

Synapse and neuron association is completed by finding the neuron labels (i.e., graph nodes) that overlap most frequently with the labeled voxels from each synapse object (i.e., graph edge).  This association is recorded via bidirectional linkages in the RAMON objects' metadata fields.  Metadata assigned to each object can be traversed server side to construct a graph \cite{Burns2013}, or the graph can be built client side at small scales.  Output graphs are converted via a web-interface to a community compatible format of choice using MROCP \cite{Mhembere}, such as GraphML.  

\subsection{Data}

Our experiments utilize a large publicly available volume of mouse somatosensory (S1) cortex, imaged using scanning electron microscopy at $3 \times 3 \times 30\;nm$ per voxel \cite{Hayworth2006a} aligned and ingested into the Open Connectome Project infrastructure.  All images were color corrected \cite{Kazhdan} and downsampled by a factor of two in the imaging plane.  The entire raw data volume is approximately 660GB.  The inscribed cube for our deployment workflow is $6000 \times 5000 \times 1850$ voxels (56 GB), or roughly 60,000 $um^3$.

\subsection{Pipeline Training}

To prepare for algorithm evaluation and testing, we first need to train a variety of algorithms used in the pipeline.  For these tasks, we select a data region separate from our evaluation region.  Our primary training region was a $1024 \times 1024 \times 100$ voxel region (known to the community as AC4).  Gold-standard annotations for both neurons and synapses exist for this volume, based on expert neuroanatomist tracings.  Our training tasks include: selecting a template for our vesicle detection module; training our deep-learning membrane classifier on 2D patches; building a random forest classifier for our synapse detection module; and training a Gala agglomerative classifier.

\subsection{Pipeline Evaluation}

\begin{figure*}[t]
\centering{
\includegraphics[height=9cm]{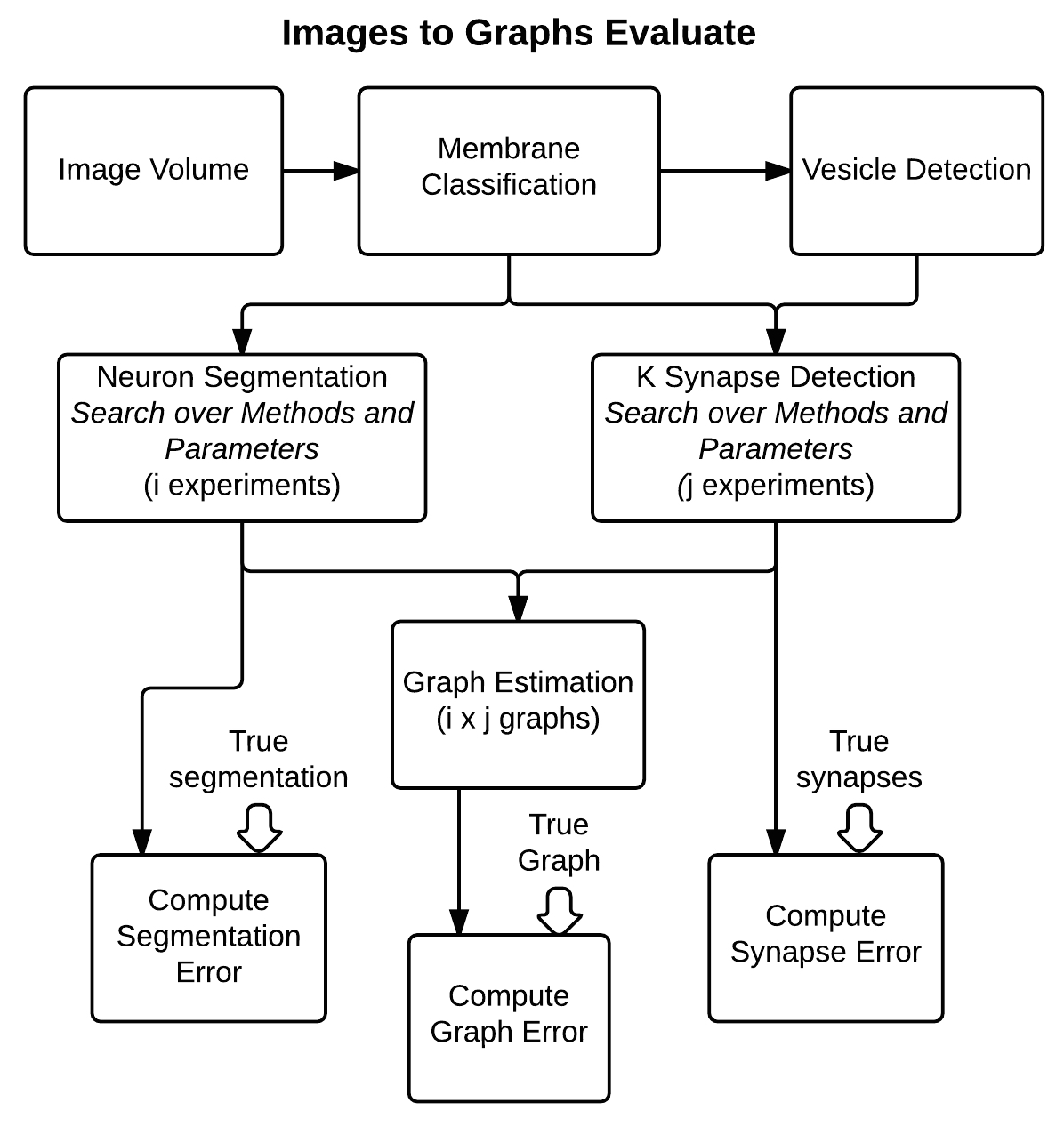}
\caption{An overall view of the Images-to-Graphs Evaluation Pipeline, beginning with image data and ending with graph creation.  Graphs are estimated and evaluated for each combination of i segmentation experiments and j synapse detection experiments.}
\label{eval-pipeline}
}
\end{figure*}

To evaluate the optimal setting for generating graphs from volumes of brain images, we constructed a fully automated pipeline to conduct a hyper-parameter search of different algorithms and their parameters and evaluate them based on community-suggested measures of synapse error, segmentation error, and our novel graph error metric (Figure~\ref{eval-pipeline}).  Other metrics can be straightforwardly added if desired.  For evaluation, we used a separate, previously unseen region ($1000 \times 1000 \times 100$ voxels), known to the community as (part of) AC3. Gold-standard annotations for both neurons and synapses exist for this volume, based on expert neuroanatomist tracings. 

\subsection{Pipeline Deployment}

\begin{figure*}[t]
\includegraphics[width=\textwidth]{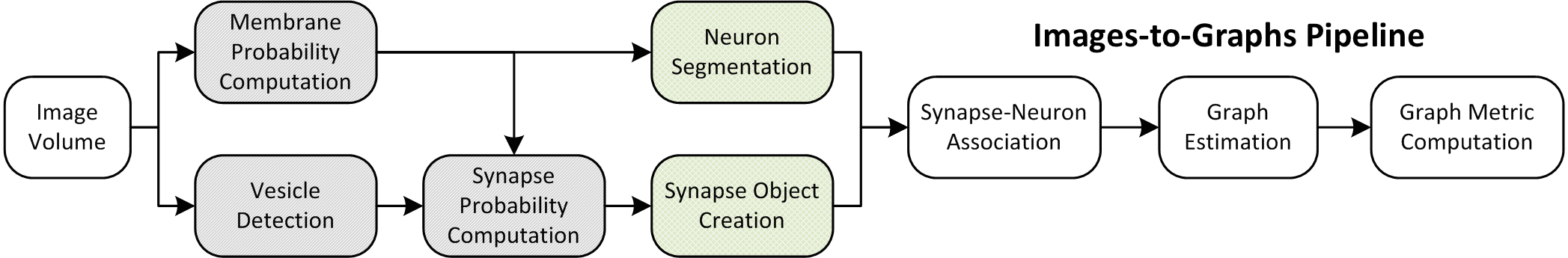}
\caption{An overall view of the Images-to-Graphs Deploy Pipeline, beginning with image data and ending with graph creation.  Modules in white are executed each time, modules that are gray (darkly shaded) are executed once and not varied in our analysis, and modules lightly shaded represent our parameter space.}
\label{deploy-pipeline}
\end{figure*}

In this pipeline, we run on a large volume for connectomic analysis (Figure~\ref{deploy-pipeline}).  Based on the classifiers created in the training workflows and the operating points found in the evaluation pipeline above, we select an operating point and deploy our end-to-end images-to-graphs pipeline as a reference implementation over a large volume (the entire inscribed dataset).  

\section{Results}

\label{sec:results}

The images-to-graphs pipeline allows us to address the question of graph quality and begin to optimize results. We take a systems view of the connectomics problem and evaluate a set of hyper-parameters (i.e., both entire algorithms and parameters within algorithms) to determine the best operating point.  In principle, parameters across all modules can be explored; we limited our experiment to variations in neuron segmentation and synapse detection methods for simplicity.

\subsection{Experiments}

We initially performed a parameter sweep to determine the best operating point for our chosen metric, and then applied those parameter settings in a deployed setting.

\subsubsection{Evaluation}
\begin{figure*}[tb]
\includegraphics[width=\textwidth]{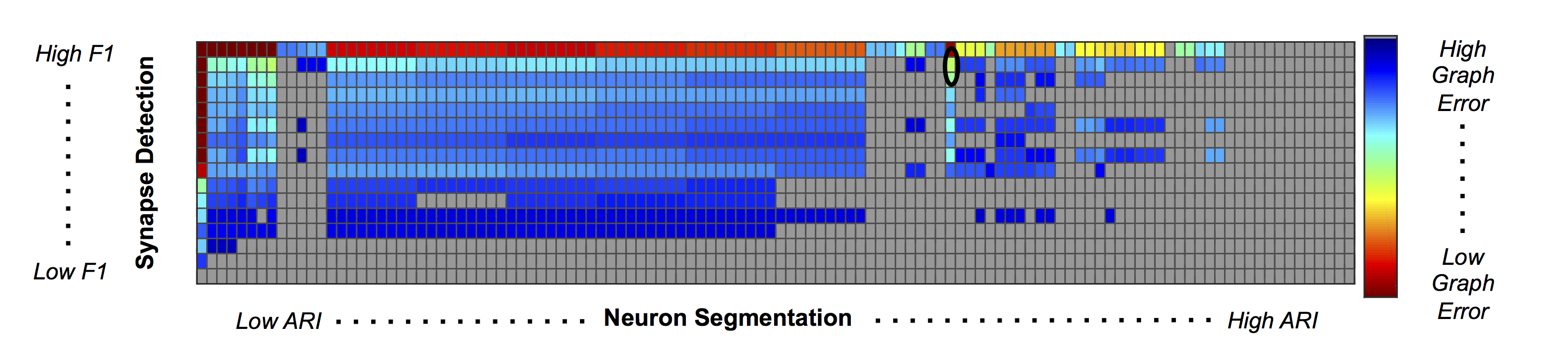}
\caption{Experimental Graph Based Error.  1856 graphs were created by combining 13 synapse detector operating points (rows) with 100 neuron segmentation operating points (columns).  The rows are ordered by synapse F1 score, and the columns by segmentation adjusted Rand index.  The first row and column represent truth, and the upper left corner of the matrix has an error of 0.  Cell color reflects graph error (clipped to show dynamic range), with a dark blue indicating lowest error and red indicating highest error.  Values shaded in gray are not significant; the selected operating point (max F1 graph score is circled in black.}
\label{graphmetrics}

\end{figure*}

We used our pipeline to examine the interaction and settings of the segmentation algorithm and the synapse detector that achieve the optimal graph f1 score.  Our evaluation varied neuron segmentation parameters (e.g., membrane strength, thresholds, number of segmentation hypotheses).  Our synapse operating points were chosen by sweeping over size and probability thresholds.  Combinations of these parameters were tested, and the results are displayed as a matrix in Figure~\ref{graphmetrics}. We examined 1856 possible graphs, requiring approximately 8,000 cluster jobs and over 3TB of data manipulation.  The entire evaluation workflow takes approximately 13 hours.

\begin{figure}[h]
\begin{center}
\includegraphics[width=11.5cm]{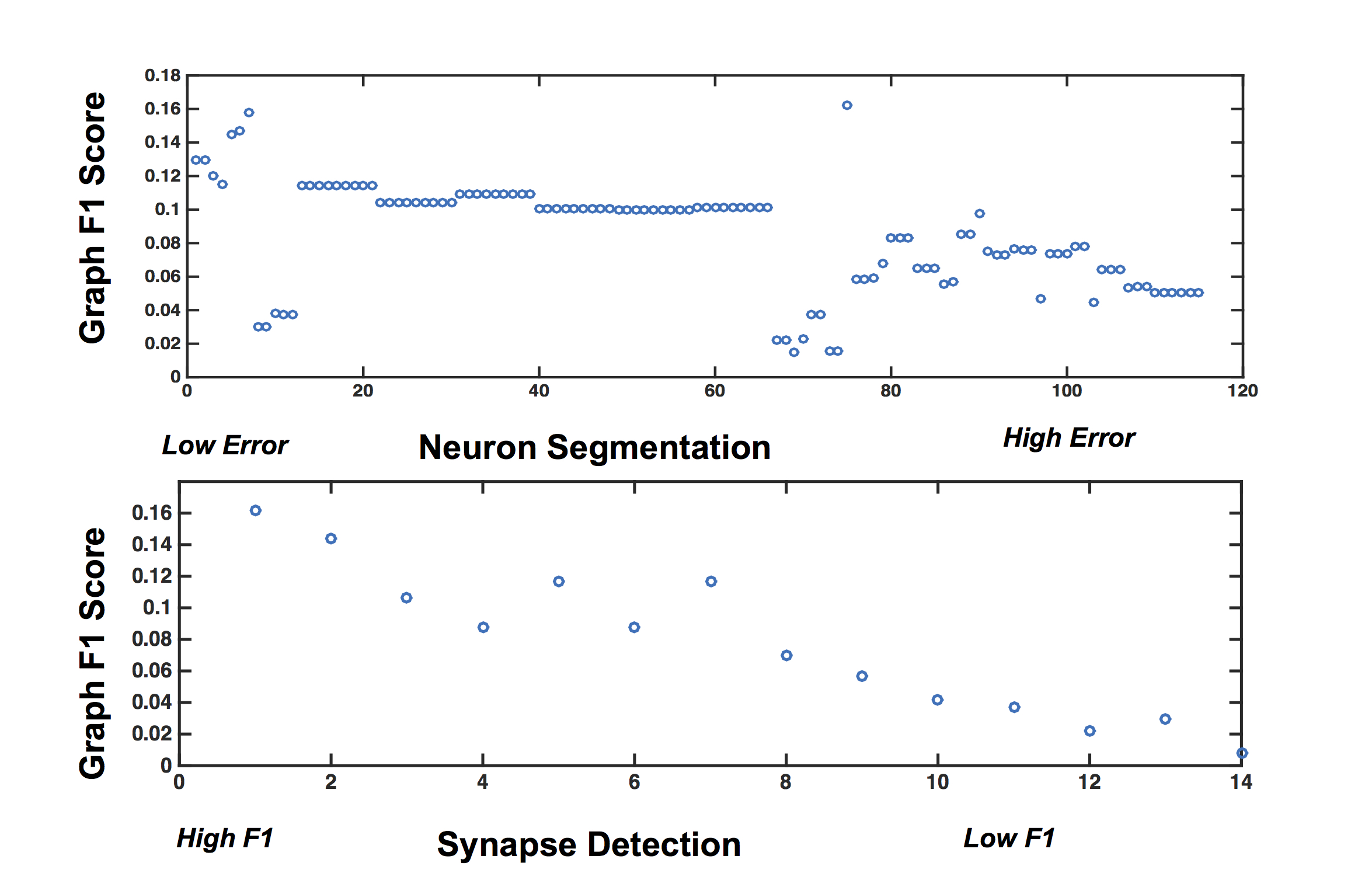}
\caption{Plots showing the variability of graph error with segmentation error (top) and synapse error (bottom), for the rows and columns associated with the best operating point.}
\label{corrplot}
\end{center}
\end{figure}

After synapses and neurons were combined to construct a graph, we evaluated the line graph error.  A permutation test was run to compute the null distribution of this test statistic.  Specifically, we calculate the graph error by uniformly sampling a random graph with the same line graph density as the observed graph for B=10,000 samples.  The p-value is then the value of the resulting cumulative distribution function, evaluated at the test-statistic value of the observed graph.  We chose a p-value significance threshold of less than 0.001; non-significant operating points are shown in gray in Figure~\ref{graphmetrics}.

Figure~\ref{graphmetrics} shows the results, in sorted synapse and segmentation error order.  Each cell in the matrix represents a single graph, and the optimal result is circled in the table.
The best result occurs at a segmentation ARI much worse than optimal, and at the maximum F1 score.  It is clear that constructing the best graph (according to our chosen metric) is more complicated than simply choosing the point with the best synapse F1 score and lowest segmentation adjusted rand index error.  Figure~\ref{corrplot} further demonstrates the non-linear relationship between graph error and intermediate measures.  By considering the overall problem context, we can select and tune the available algorithms to determine the best result (i.e., operating point) for a given task, such as motif finding.   The optimal graph was computed using the Gala segmentation algorithm with an agglomeration threshold of 0.8; the synapse detection probabilities were thresholded at 0.95, and a connected component analysis was used to form the final synapse objects.  Objects with a size greater than 5000 pixels in 2D or less than 1000 voxels in 3D were removed to reduce erroneous detections.  The optimal F1 score was 0.16, indicating that significant improvement is needed.

\subsubsection{Deployment}
The deployment workflow provides a capability demonstration and produced 12234 neurons with non-zero degree and 11489 synaptic connections in a volume of $\approx 60,000$ cubic microns.  Total runtime on 100 cores was about 39 hours, dominated by the block merging step, which is currently performed on each seam serially.  Membrane computation currently takes an additional 3 weeks on our small GPU cluster; this operation is embarassingly parallel and recent advances suggest methods to dramatically speed up this step \cite{Masci2013}.

\section{Discussion}

\label{sec:discussion}

We have demonstrated the first framework for estimating brain graphs at scale using automated methods.  We recast the problem as a graph estimation task and consider algorithm selection and parameter tuning to optimize this objective by leveraging a novel graph error metric.  This work provides a scaffolding for researchers to develop and evaluate methods for the overall objective of interest. 

We evaluate our pipeline across a set of parameters and modules, leveraging a combination of published methods and novel algorithms.  Additional insights may be gained at larger scales and through additional optimization.  Although our error metric currently considers only  binary, unweighted graphs, there are opportunities to extend this to apply to attributed graphs, as well as to weight the metric by error types (e.g., number of false positives or false negatives). 

Automated results do not need to be perfect to draw statistical conclusions, and errorful graphs may be used as the basis for inference and exploitation of ``big neuroscience" challenges \cite{Priebe2012}.  Bias in errors is another important factor in constructing exploitable graphs that is not fully explored in this manuscript. With the ability to efficiently compare combinations of different algorithms and operating points, we can begin to answer the question of graph quality and how to optimize the results.  Having the ability to examine the process from an end-to-end systems approach will enable rapid improvement in graph quality.   The infrastructure demonstrated in this work provides a community test-bed for further exploration and at scale computation.  Although this manuscript focuses exclusively on electron micrographs, our framework is extensible to many other modalities, including Array Tomography, CLARITY, and two-photon calcium imaging data.

\section*{Data Sharing}

All code, algorithms, documentation and data products are open source and released under an Apache 2.0 license.  These are linked to from the Images-to-Graphs project webpage at:  \url{i2g.io}

\section*{Disclosure/Conflict-of-Interest Statement}

The authors declare that the research was conducted in the absence of any commercial or financial relationships that could be construed as a potential conflict of interest.

\section*{Author Contributions}

WGR, DMK, RJV, MAC, and GDH designed and conceived of the experiments and provided technical expertise. Graph metrics and statistical analysis was performed by WGR, JTV, and CEP.  WGR, DMK, and MP designed and implemented computer vision algorithms.  PM, KL, RB, JTV, WGR, and DMK built the Open Connectome Project and related infrastructure.  WGR wrote the manuscript and ran the final experiments, with inputs from all authors.

\section*{Acknowledgments}
The authors wish to thank Dennis Lucarelli and Aurora Schmidt for their help in algorithm development and Greg Kiar for helpful feedback on manuscript drafts.  We also thank Bobby Kasthuri, Daniel Berger, and Jeff Lichtman for providing electron microscopy data and truth labels.

\section*{Funding} This work is partially supported by JHU Applied Physics Laboratory Internal Research and Development funds and by NIH NIBIB 1RO1EB016411-01.  

\bibliographystyle{ieeetr} 

\bibliography{i2g.bib}

\end{document}